\documentclass[prb,preprint,superscriptaddress,preprintnumbers,a4paper,amsmath,amssymb,showpacs,floatfix]{revtex4}
\usepackage{graphicx}
\usepackage{color}\color[rgb]{0.000,0.000,0.000} 
\usepackage{amssymb} 
\usepackage{amsmath} 
\usepackage{multirow}

\begin{document}

\newcommand{\ba}{Ba$_{3}$NaRu$_2$O$_9$}
\newcommand{\five}{Ru$^{5+}_2$O$_9$}
\newcommand{\six}{Ru$^{6+}_2$O$_9$}
\newcommand{\te}{$_{t}$}
\newcommand{\ot}{$_{o}$}
\newcommand{\Tc}{T$_{C}$}
\newcommand{\Ts}{T$_{s}$}
\newcommand{\Tn}{$T_\mathrm{N}$}
\newcommand{\MuB}{$\mu_\mathrm{B}$}
\newcommand{\mn}{MnS$_{2}$}

\title{Giant pressure-induced volume collapse in the pyrite mineral \mn}
\author{Simon A. J. Kimber \footnote{These authors contributed equally to this work.\\
CLASSIFICATION: Physical Sciences, Earth, Atmosphere and planetary sciences.}}
\email[Email of corresponding author:]{kimber@esrf.fr}
\affiliation{European Synchrotron Radiation Facility (ESRF), 6 rue Jules Horowitz, BP 220, 38043  Grenoble Cedex 9, France;}
\author{Ashkan Salamat $^{*}$}
\email[Email of corresponding author:]{salamat@physics.harvard.edu}
\affiliation{European Synchrotron Radiation Facility (ESRF), 6 rue Jules Horowitz, BP 220, 38043  Grenoble Cedex 9, France;}
\affiliation{Lyman Laboratory of Physics, Harvard University, Cambridge, MA 012138, USA.}
\author{Shaun R. Evans}
\affiliation{Universit\"at Bern, Departement f\"ur Chemie und Biochemie, Freiestrasse 3, CH-3012 Bern, Switzerland;}
\author{Harald O. Jeschke $^{*}$}
\email[Email of corresponding author:]{jeschke@th.physik.uni-frankfurt.de}
\affiliation{Institut f\"ur Theoretische Physik, Goethe-Universit\"at Frankfurt, Max-von-Laue-Stra\ss e 1, 60438 Frankfurt am Main, Germany;}
\author{Kaliappan Muthukumar}
\affiliation{Institut f\"ur Theoretische Physik, Goethe-Universit\"at Frankfurt, Max-von-Laue-Stra\ss e 1, 60438 Frankfurt am Main, Germany;}
\author{Milan Tomi\'c}
\affiliation{Institut f\"ur Theoretische Physik, Goethe-Universit\"at Frankfurt, Max-von-Laue-Stra\ss e 1, 60438 Frankfurt am Main, Germany;}
\author{Francesc Salvat-Pujol}
\affiliation{Institut f\"ur Theoretische Physik, Goethe-Universit\"at Frankfurt, Max-von-Laue-Stra\ss e 1, 60438 Frankfurt am Main, Germany;}
\author{Roser Valent\'i}
\affiliation{Institut f\"ur Theoretische Physik, Goethe-Universit\"at Frankfurt, Max-von-Laue-Stra\ss e 1, 60438 Frankfurt am Main, Germany;}
\author{Maria V. Kaisheva}
\affiliation{Formerly at the School of Chemistry, Joseph Black Building, King's Buildings, EH9 3JJ, Edinburgh, United Kingdom;}
\author{Ivo Zizak}
\affiliation{Helmholtz-Zentrum Berlin f\"ur Materialien und Energie (HZB), BESSY-II, Wilhelm Conrad R\"ontgen Campus, Albert-Einstein-Str. 15, 12489 Berlin, Germany;}
\author{Tapan Chatterji}
\affiliation{Institut Max von Laue-Paul Langevin, 6 rue Jules Horowitz, BP 156, F-38042, Grenoble Cedex 9, France.}
\date{\today}
\begin{abstract}
\textbf{Dramatic volume collapses under pressure are fundamental to geochemistry\cite{Cohen,Lin}~and of increasing importance to fields as diverse as hydrogen storage\cite{Hydrogen}~and high temperature superconductivity\cite{Andreas}. In transition metal materials, collapses are usually driven by so-called 'spin state' transitions- the interplay between the single-ion crystal field and the size of the magnetic moment\cite{Cohen,Lin}. Here we show that the classical S= $\frac{5}{2}$~mineral Hauerite\cite{Hastings} (MnS$_{2}$) undergoes an unprecedented\cite{Cohen,Lin,Cerium,Kunes,Chatt,Chatt2,PNAS}~($\Delta V\sim$ 22 \%) collapse driven by a conceptually new magnetic mechanism. Using synchrotron x-ray diffraction we show that cold compression induces the formation of a disordered intermediate. However, using an evolutionary algorithm\cite{USPEX1,USPEX2}~we predict a new structure with edge-sharing chains. This is confirmed as the thermodynamic ground state using in-situ laser heating. We show that magnetism is globally absent in the new phase, as low-spin quantum S= $\frac{1}{2}$~moments are quenched by dimerisation\cite{MgTi2O4}. Our results show how the emergence of metal-metal bonding can stabilise giant spin-lattice coupling in Earth's minerals.}\\

\noindent SIGNIFICANCE STATEMENT: Minerals which contain magnetic metals may collapse under the pressures found in the Earth's mantle. These so-called spin-state transitions are due to the reduction of the magnetic moment associated with each metal atom. Here we report the discovery of a giant volume collapse in the mineral Hauerite (MnS$_{2}$) under pressure. Instead of a change in the single-ion magnetic moments, this is driven by the Mn$^{2+}$~cations spontaneously forming pairs, or dimers. In contrast to the magnetic, unpaired electrons found at ambient pressure, the dense new phase thus contains an ordered arrangement of chemical bonds which are globally non-magnetic. This 'squeezing out' of magnetism is shown to stabilize the huge increase in density.
\end{abstract}
\maketitle
Magnetism plays an unsurprising role in determining the properties of the minerals which make up Earth's crust and mantle. This is due to the ubiquity of transition metals (notably Fe) with unpaired electrons. Under geological pressures, these may undergo abrupt volume collapses due to spin-state transitions. These occur when the crystal field splitting $\Delta$, competes with Hund's rule and electronic Coulomb repulsion to determine the size of the magnetic moment. High pressure can thus make a high-spin, e.g. S= $\frac{5}{2}$~state unstable with respect to a low-spin state\cite{Cohen,Kunes}, as $\Delta$~increases as the metal-ligand bond distances decrease. For simple minerals\cite{Hematite,Siderite},  volume collapses of ~$\Delta V\simeq$ 5 \%~are regarded as notable\cite{PNAS,Hematite,Siderite}. Understanding the mechanisms\cite{Kunes}~of  such magnetically driven transitions, and predicting the high-pressure structures, is ultimately of great importance for modelling the mantle in particular\\
The simple pyrite structured mineral Hauerite (\mn) was reported to undergo a pressure-induced transition nearly thirty years ago\cite{Chatt,Chatt2}. The pyrite structure\cite{Vesta}~(Fig. 1a) is cubic, with molecular disulphide S$_{2}^{2-}$~groups, which octahedrally coordinate an \textit{fcc}~lattice of Mn$^{2+}$~sites. Under ambient conditions, a high-spin \textit{t$_{2g}^{3}$}\textit{e$_{g}^{2}$}~S = $\frac{5}{2}$~moment is stabilised by Hund's rule coupling, and long-range magnetic order results\cite{Hastings}. Early work on \mn, which used laboratory energy dispersive diffraction\cite{Chatt}, detected a structural change at 11 GPa. These results were interpreted as a spin-state transition to a \textit{t$_{2g}^{5}$}\textit{e$_{g}^{0}$}~S = $\frac{1}{2}$~ state in the marcasite structure. This was supported by later calculations\cite{Persson}. However, when the experiment was repeated using synchrotron radiation\cite{Chatt2}, an unidentified disordered phase was observed. Together with the large estimated volume collapse ($\sim$15 \%), this little-remarked upon finding was one of the motivations for the present work. \\
We used the same natural sample as previous investigations\cite{Chatt,Chatt2}, and confirmed its purity by diffraction and x-ray fluorescence measurements\cite{Sole}~(\textbf{S1}). Pressure was applied using gas-loaded diamond anvil cells. Up to 11.7 GPa, x-ray diffraction (Fig. 1b) shows that the cubic \textit{Pa-3}~pyrite structure is preserved, with no reduction in crystallinity and a low bulk modulus of B = 65.9(3) and B' = 5.1(2) GPa. Upon exceeding 11.85 GPa,  a complete switch occurs. The resolution-limited pyrite peaks are replaced by a series of broad maxima, which render structural analysis impossible. The diffraction pattern continues to evolve up to 29.1 GPa, suggesting a substantial region of phase co-existence. Due to the poor quality of the diffraction data on cold compression, we undertook structure searching using density functional theory (\textsc{dft}) methods. Structure relaxation using \textsc{vasp} yielded low-spin metallic pyrite structures as reported previously\cite{Persson}. However, this could not explain the complex diffraction patterns observed. We therefore used\cite{USPEX1,USPEX2}~the Universal Structure Predictor: Evolutionary algorithm package (\textsc{uspex}) to generate and compare a large number (ca. 2800) of possible structures. Candidates were locally relaxed using the Vienna ab initio simulation package (\textsc{vasp}) implementation\cite{VASP, VASP2, VASP3, Bloechl} of \textsc{dft}~within the generalised gradient approximation\cite{GGA}. This procedure\cite{Muthukumar}~yielded the (dynamically stable, \textbf{S2}) arsenopyrite structure shown in Fig. 2a. The disulphide dimers are maintained, although the connectivity of the MnS$_{6}$~octahedra is dramatically changed with the formation of one-dimensional edge-sharing chains. This structure search was completely \textit{ab-initio}, with no experimental input.\\ 	
	To experimentally validate these results, we applied \textit{in-situ}~CO$_{2}$ laser heating to the disordered state at 20 GPa. In a matter of seconds (\textbf{S3}), sharp peaks appeared, persisting to our maximum temperature of $\sim$1800 K. When the laser was switched off, we obtained a phase-pure diffraction pattern. This was indexed with a primitive monoclinic cell, equivalent to that given by our \textsc{uspex}~calculations, in space group \textit{P2$_{1}$/c} with $\beta$$\approx$111 $^{\circ}$, and Rietveld refinement of the arsenopyrite structure proceeded smoothly (Fig. 2b).  A notable distortion of the edge-sharing chains was found in the refined and calculated structures (Fig. 2c). Alternate manganese sites are displaced to form dimer pairs, with a large variation in Mn-Mn distances from 2.72 to 3.42 \AA.\\
	The post-laser heating arsenopyrite phase was stable on decompression (\textbf{S4}) to 6 GPa, before a mixed phase region, then complete reversion to pyrite was observed.  Comparison with the pyrite cell volume at 11 GPa showed a 22 \% reduction in volume (Fig. 3a). This is by far the largest reversible volume collapse reported for any transition metal oxide or sulfide. Upon further decompression into the region of coexistence with pyrite below 6 GPa, the volume difference peaks at around 25 \%. At ambient conditions, the recovered pyrite lattice parameter was identical to that before compression. We confirmed the stability of the pyrite and arsenopyrite phases independently, by calculating their relative enthalpies. A crossover at 10 GPa is found (Fig. 3b),   in excellent agreement with experiment.\\
A change in the spin-state of \mn~should be reflected in the bond lengths. However, extracting this information from experiment alone is difficult, due to the low-symmetry and texture effects on decompression (\textbf{S5}). We therefore calculated the pressure dependence of the Mn-Mn and Mn-S distances in both the pyrite and arsenopyrite structures. As shown in Fig. 3c and 3d, a large reduction in the Mn-Mn  distances is found at the pyrite$\rightarrow$arsenopyrite transition. We also reproduced the dimerisation in the edge-sharing chains. This can be seen to be enhanced under pressure, with a marked increase above 16 GPa. When we compared these results with those extracted from Rietveld refinement\cite{GSAS} against the experimental data, we found near-perfect agreement. Similar excellent agreement can be seen for the important $<$Mn-S$>$ distance. This should be expanded when \textit{e$_{g}$} electron density is present (in the pyrite phase), and contract in a putative \textit{t$_{2g}^{5}$}\textit{e$_{g}^{0}$} low-spin state. Exactly this change in the $<$Mn-S$>$~distances is found in theory and experiment with a contraction from 2.45$\rightarrow$2.24 \AA.\\ 
We used a range of \textsc{dft}~methods to analyse the electronic structures of both phases of \mn~in more detail\cite{Wien2K,FPLO}. At ambient pressure, our results for the pyrite structure are similar to those previously reported\cite{Persson}. Type-III antiferromagnetic order\cite{Hastings}~is lowest in energy and the calculated band gap is 1.3 eV. A large Mn spin-polarisation is evident in the electronic density of states (Fig. 4a), consistent with an S=$\frac{5}{2}$~moment. For the arsenopyrite structure (Fig. 4b), a slightly reduced band gap (1 eV) is obtained, however, the highlighted density of states for Mn in the spin-up and spin-down channels is almost identical, showing that the moment has collapsed. The full pressure dependence of the magnetic moment is shown in Fig. 4c. This unambiguously shows that the volume collapse in \mn~is of magnetic origin. In addition, the correlation between the Mn-Mn dimerisation, moment quenching and band gap (\textbf{S6}) is demonstrated by the further drop at 16 GPa.\\
The above results show that the size of the volume collapse in \mn~not only exceeds that of previously reported electronic effects, including the benchmark example of Cerium\cite{Cerium}, but is comparable to the volume drop found when open framework materials chemically decompose\cite{Arora}. This explains why a kinetically-hindered intermediate conceals\cite{Chatt2}~the thermodynamic ground state on isothermal compression. Unlike other classic examples of pressure-induced spin-state transitions\cite{Kunes}, there is no ambiguity about the role of band structure effects, as the arsenopyrite phase of \mn~is insulating (Fig. 4b). However, the dimerised superstructure which emerges deserves further comment. Isoelectronic CoSb$_{2}$ also crystallises with the arsenopyrite structure at ambient pressure, however, on warming above 650 K, the dimer distortion is lost and a magnetic susceptibility develops\cite{Siegrist}. Although this has been discussed within the framework of a Peierls-type distortion\cite{Wijeyesekera}, our calculations show that \mn~is actually better described as a 3D-network of dimers. The hopping parameters, \textit{t}, extracted from our electronic structure calculations (\textbf{S7}) are comparable ($\lesssim$0.15 eV) in all crystallographic directions, with the notable exception of the intradimer hopping (0.37 eV). The ordered pattern of short Mn-Mn distances thus corresponds to formation of a valence bond solid\cite{MgTi2O4}. This ground state corresponds to S=0 Mn-Mn dimers formed by S=$\frac{1}{2}$~Mn ions. Such a ground state is disfavoured by larger atomic spins. In CrSb$_{2}$, which is S=1, the edge-shared chains remain undistorted, instead showing one-dimensional spin wave excitations\cite{Stone}. Valence bond solid formation in \mn~is therefore strongly linked to the quantum nature of the S=$\frac{1}{2}$~low-spin state of Mn that is favoured by high-pressure conditions.\\
We believe that the giant volume collapse, as well as the total loss of magnetic moments without metallisation are therefore the result of a quantitatively new mechanism, which goes beyond the simple single-ion paradigm for transition metal materials\cite{Tokura}. While crystal field changes are important, the ultimate stabilising factor is the surprising formation of dimer interactions in a continuous network solid. The resulting metal-metal bond enthalpy presumably outweighs the cost of the huge increase in density in the arsenopyrite phase. Finally, we note that \mn~ is geologically scarce, however, the mechanism which stabilises the volume collapse is valid for more abundant minerals containing isoelectronic cations like Fe$^{3+}$.\\

\pagebreak
\noindent\textbf{Acknowledgements}\\
We thank the ESRF, Grenoble, and BESSY-II, Berlin for access to synchrotron facilities. We acknowledge M. Hanfland for assistance on ID9A. We thank E. Gregoryanz, J.S. Loveday, W. Crichton and P. Bouvier for useful discussions. HOJ, KM, MT, FSP and RV would like to thank the Deutsche Forschungsgemeinschaft for financial support through grant SFB/TR49 and the Beilstein Institut through Nanobic. We also gratefully acknowledge the centre for scientific computing (CSC, LOEWE-CSC) in Frankfurt for computing facilities.\\

\noindent\textbf{Author Contributions}\\
The sample and inspiration for this work were provided by TC and diffraction experiments were performed by AS, SAJK and SE. Data were analysed by SAJK, AS and SE. XRF analysis was performed by MVK and IZ. Calculations were performed by HOJ, KM, MT, FSP and RV. SAJK wrote the paper with assistance from AS and the other authors.\\

\noindent\textbf{Methods}\\
A natural single crystal was obtained from the collection of the late Prof. H-G von Schnering.  This sample is believed to have been acquired from the Mineralogisches Museum M\"unster ca. 1982. A small piece was cracked off and ground into a fine powder for the measurements reported here. The sample was characterised by high resolution synchrotron powder x-ray diffraction using ID31 (E$_{i}$ = 31 KeV) at the ESRF, Grenoble, and by x-ray fluorescence measurements at the MySpot beamline of the BESSY II synchrotron, Berlin. All diffraction data were analysed using the GSAS suite of software. The XRF spectra were fitted using PyMca\cite{Sole}.

High pressure diffraction data were acquired using beam lines ID9A (E$_{i}$= 30 KeV) and ID27 (E$_{i}$ = 33 KeV) at the ESRF. Mar555 and Mar135 CCD detectors were used respectively. The high pressure experiments were carried out in membrane driven diamond anvil cells using 300 micron diamonds and rhenium gaskets. For the ambient temperature compression runs helium was used as the pressure transmitting medium whilst for the laser heating experiments at high pressure neon was used, also acting as a thermal insulator. The pressure was calibrated using the re-calibrated ruby fluorescence scale\cite{Ruby}. Laser heating was carried out in situ at ID27 using a CO$_{2}$ laser ($\lambda$ = 10.6 $\mu$m) and the synthesis processed monitored with x-rays.  Temperature was calculated from the thermal emission measurements collected from the hotspot of the CO$_{2}$~laser irradiation and fitted to either a Planck or Wein function. 

We employed evolutionary algorithms (\textsc{uspex}) developed by A. Oganov\cite{USPEX1,USPEX2}~to perform an unbiased search for the high pressure structure of \mn. ÊEach generation contained between 10 and 25 structures and the first generation was always produced randomly. ÊAll structures were locally optimised during structure search using \textsc{dft}~with the projector augmented wave\cite{Bloechl}~(\textsc{paw}) as implemented\cite{VASP,VASP2,VASP3}~in \textsc{vasp}. The generalised gradient approximation\cite{GGA}~(\textsc{gga}) in the parametrisation of Perdew, Burke and Ernzerhof was used as approximation for the exchange and correlation functional.

Based on the P=0 GPa pyrite and P=20 GPa arsenopyrite structures of \mn, compression and decompression calculations were performed using the \textsc{paw}~basis\cite{Bloechl}~as implemented in \textsc{vasp}. The structural optimisations were done using a GGA+U functional for the Mn \textit{3d}~states with U=3 eV and J=1 eV and type 3 antiferromagnetic order. We checked for other magnetic orderings but found them to be always higher in energy. Relaxations under constant pressure were performed using the conjugate gradient algorithm. Relaxations were performed in three subsequent passes, and for all relaxations a 300 eV plane wave cutoff was used in combination with a 6x6x4 Monkhorst-Pack k-mesh.

We finally analysed the electronic structure using two full potential all electron \textsc{dft}~methods. ÊWe performed calculations of the total energy, of the magnetic moments and of the density of states for the relaxed pyrite and arsenopyrite structures using the linear augmented-plane-wave basis set as implemented\cite{Wien2K}~in the code Wien2K and using the full potential local orbital basis\cite{FPLO}~(\textsc{fplo}). We adopted the \textsc{gga}~approximation to the exchange-correlation functional, employing \textsc{gga+u}~functionals for the Mn \textit{3d}~orbitals with U=3 eV and J=1 eV in the fully localised limit, and type-III antiferromagnetic order. The Wien2k calculations were carried out with 200 k points and RK$_{max}$ =9, resulting in converged energies within a tolerance of 0.05 eV. The \textsc{fplo}~calculations were performed on a 10x10x10 k-mesh with the same \textsc{gga+u}~functional and magnetic order.

\begin{figure}[tb!]
\begin{center}
\includegraphics[scale=0.13]{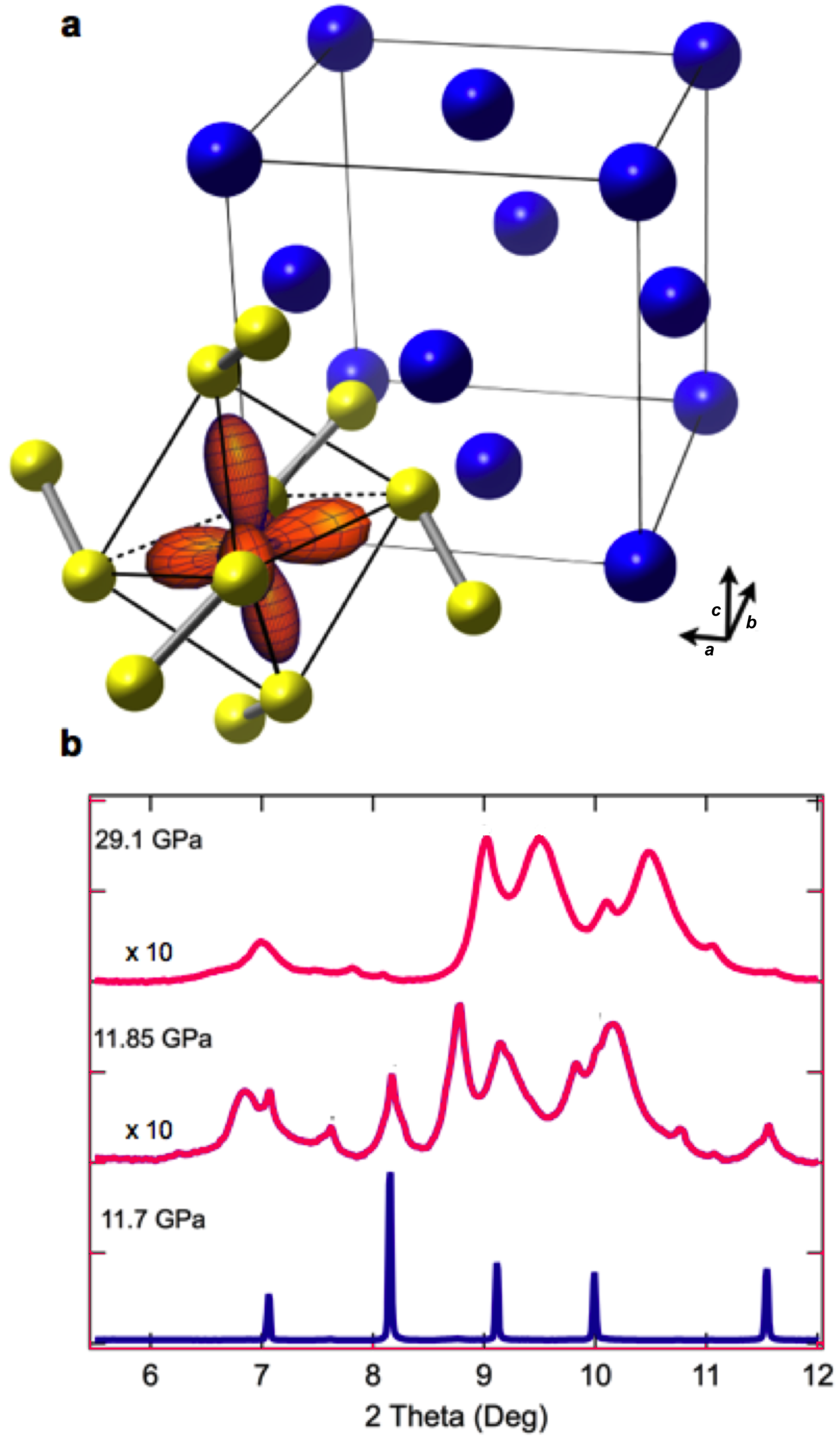}
\caption{\textbf{Ambient pressure structure of \mn~and pressure evolution of diffraction profile.} \textbf{a}, Perspective view of the structure of pyrite \mn, highlighting the \textit{fcc}~lattice of Mn sites (blue spheres) and the disulphide units (yellow dumbbells). The octahedral coordination and the valence \textit{e$_{g}$} orbitals are highlighted for one site.\textbf{b},  Pressure evolution of the x-ray diffraction profile of \mn~at ambient temperature showing the collapse into a poorly-crystalline phase. Note that the data sets above 11.7 GPa have been multiplied by a factor of 10 for clarity.}
\label{Fig1}
\end{center}
\end{figure}

\begin{figure}
\begin{center}
\includegraphics[scale=0.55]{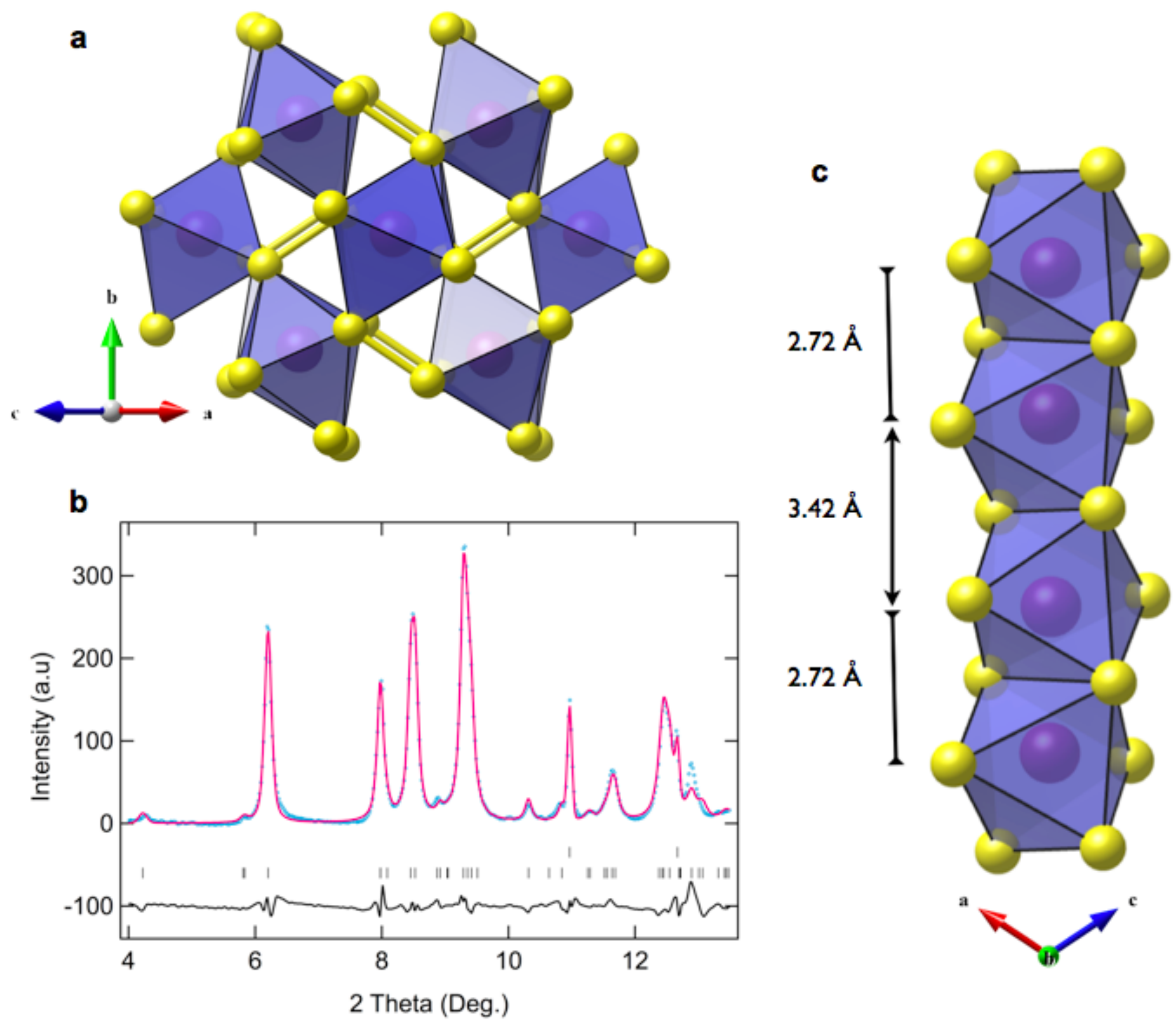}
\caption{\textbf{High-pressure structure and powder diffraction profile of \mn.} \textbf{a},  Projection of the arsenopyrite high pressure structure of \mn, as predicted by USPEX and as refined against the diffraction profile shown below. This view point is down the chains of edge-sharing MnS$_{6}$ octahedra and the arrows indicate the direction of the cell axes. \textbf{b}, Observed, calculated and difference profiles from the Rietveld fit to the x-ray diffraction pattern of \textit{P2$_{1}$/c} \mn~at 20 GPa post-laser heating. The bottom row of tick markers shows the arsenopyrite peak positions and the top row shows peak positions for the Neon pressure medium.  The refinement converged with \textit{w$_{rp}$} = 0.0985, \textit{r$_{p}$} = 0.0706 and gave cell parameters \textbf{\textit{a}} = 5.4642(6), \textbf{\textit{b}}  = 5.3742(5), \textbf{\textit{c}}  = 5.4203(7) \AA \ and $\beta$ = 111.93(2) $^{\circ}$. See tables 1-2 in the SI for more details. \textbf{c}, View of the edge-sharing chains in the refined arsenopyrite structure showing the Mn-Mn dimerisation. This view point is perpendicular to that shown in panel \textbf{a}, and the arrows are as before.}
\label{Fig1}
\end{center}
\end{figure}

\begin{figure*}
\begin{center}
\includegraphics[scale=0.48]{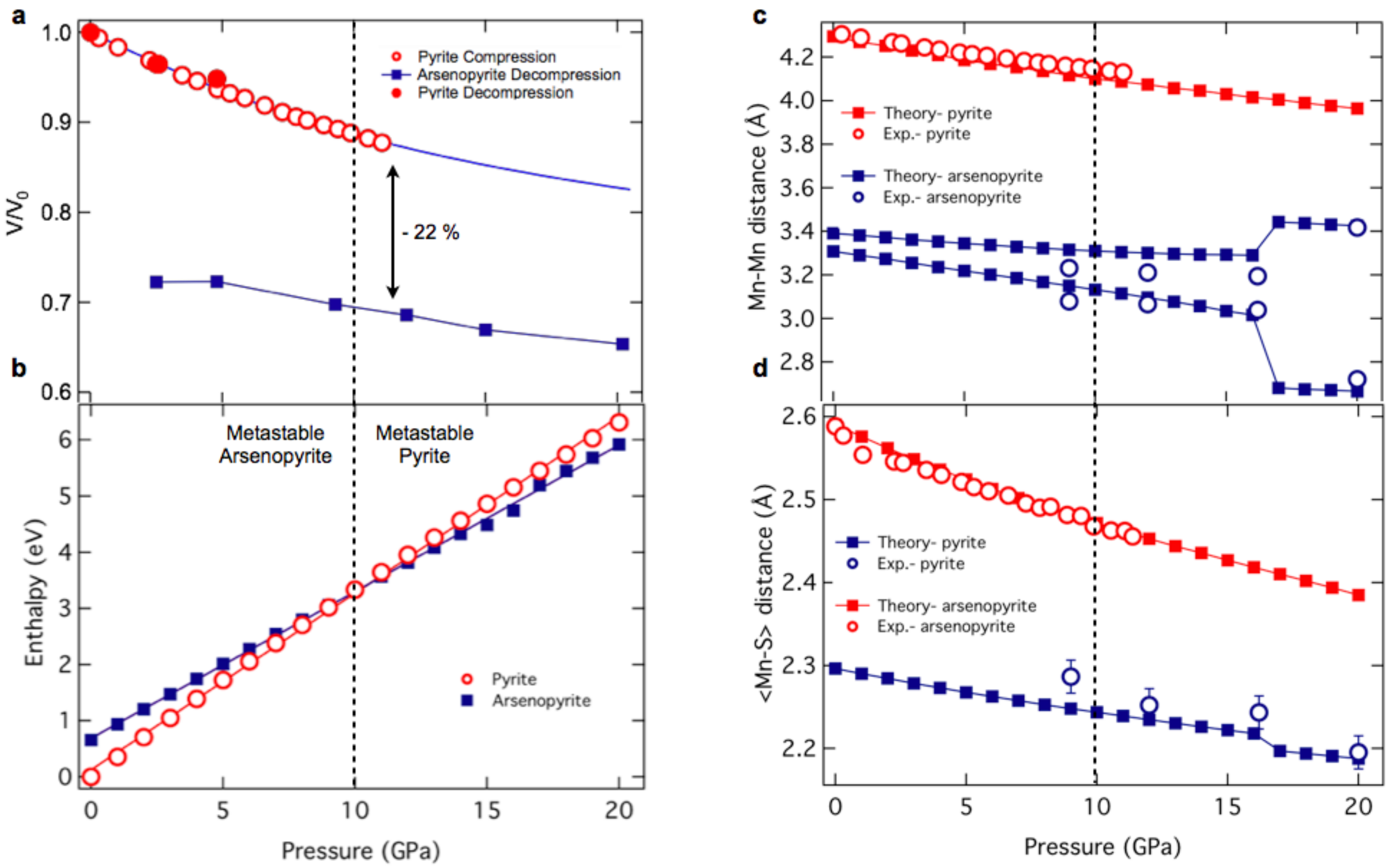}
\caption{\textbf{Pressure evolution of the unit cell volume, enthalpy and principal bond distances in \mn.} \textbf{a}, Evolution of the refined unit cell volumes (symbols) of the pyrite and arsenopyrite phases. A fit of a Birch-Murnaghan equation of state to the former is shown as a blue line. \textbf{b}, Pressure dependence of the calculated enthalpy for the  pyrite and arsenopyrite phases, showing the metastable regions for both. \textbf{c}, Pressure dependence of the Mn-Mn distances for \mn, as extracted from our structure optimisations (filled symbols) and Rietveld refinements (open symbols). \textbf{d}, Pressure dependence of the $<$Mn-S$>$~distance for  both phases of \mn. The symbols are as above.}
\label{Fig1}
\end{center}
\end{figure*}
\begin{figure*}
\begin{center}
\includegraphics[scale=0.45]{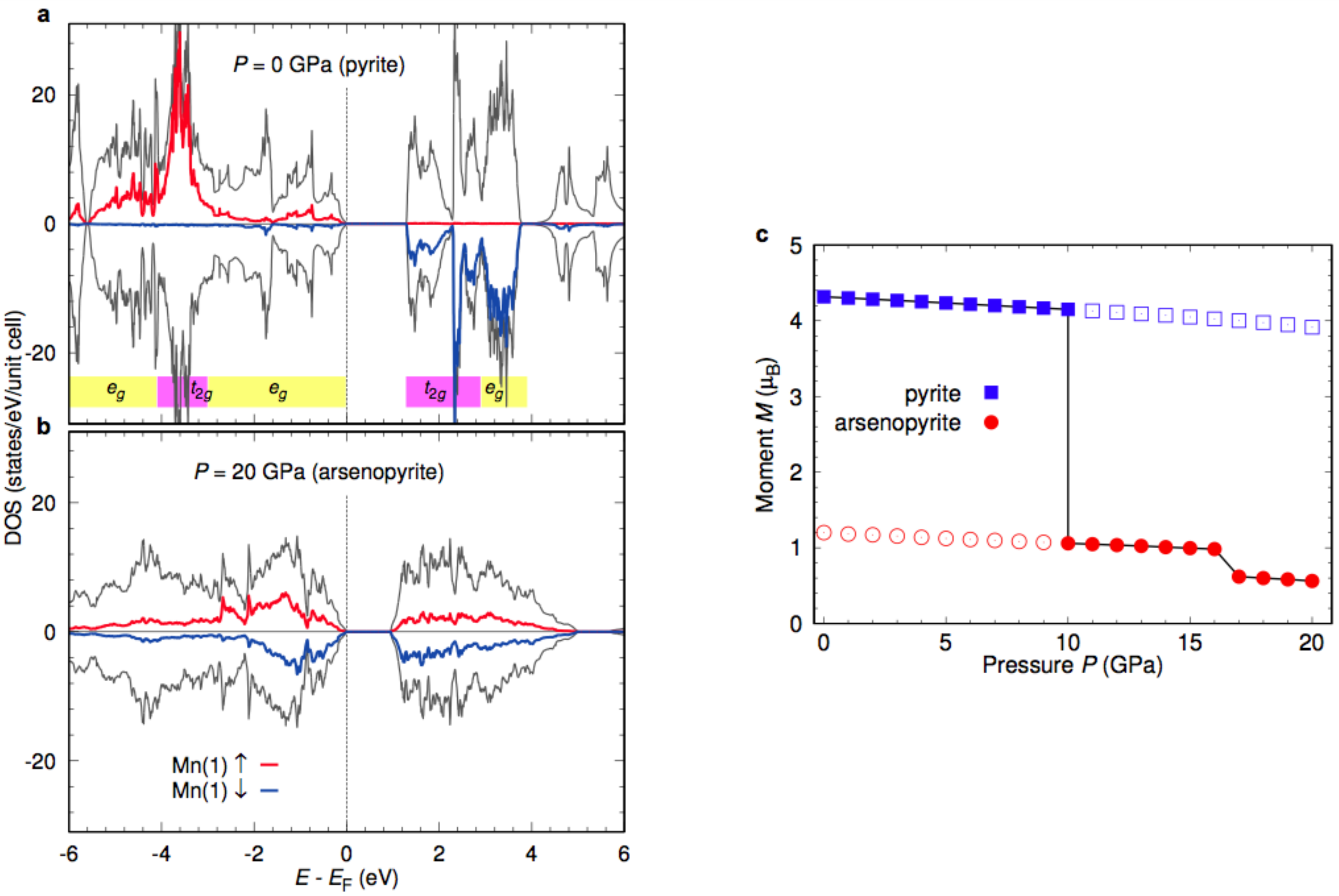}
\caption{\textbf{Pressure dependence of the electronic structure and magnetic moment of \mn.} \textbf{a}, Spin-polarised electronic density of states for pyrite structured \mn~at 0 GPa. The contributions of Mn \textit{t$_{2g}$}~and \textit{e$_{g}$}~states are highlighted and the black line is the total d.o.s.  \textbf{b}, Spin-polarised electronic density of states for arsenopyrite structured \mn~at 20 GPa. Note the near absence of spin polarisation (compare the red and blue lines).\textbf{c}, Pressure evolution of the magnetic moment in \mn~as calculated by DFT methods. Clear anomalies can be seen at the volume collapse at 11 GPa and at 16 GPa, where the Mn-Mn dimerisation is enhanced.  }
\label{Fig1}
\end{center}
\end{figure*}

\begin{figure}
\begin{center}
\includegraphics[scale=0.65]{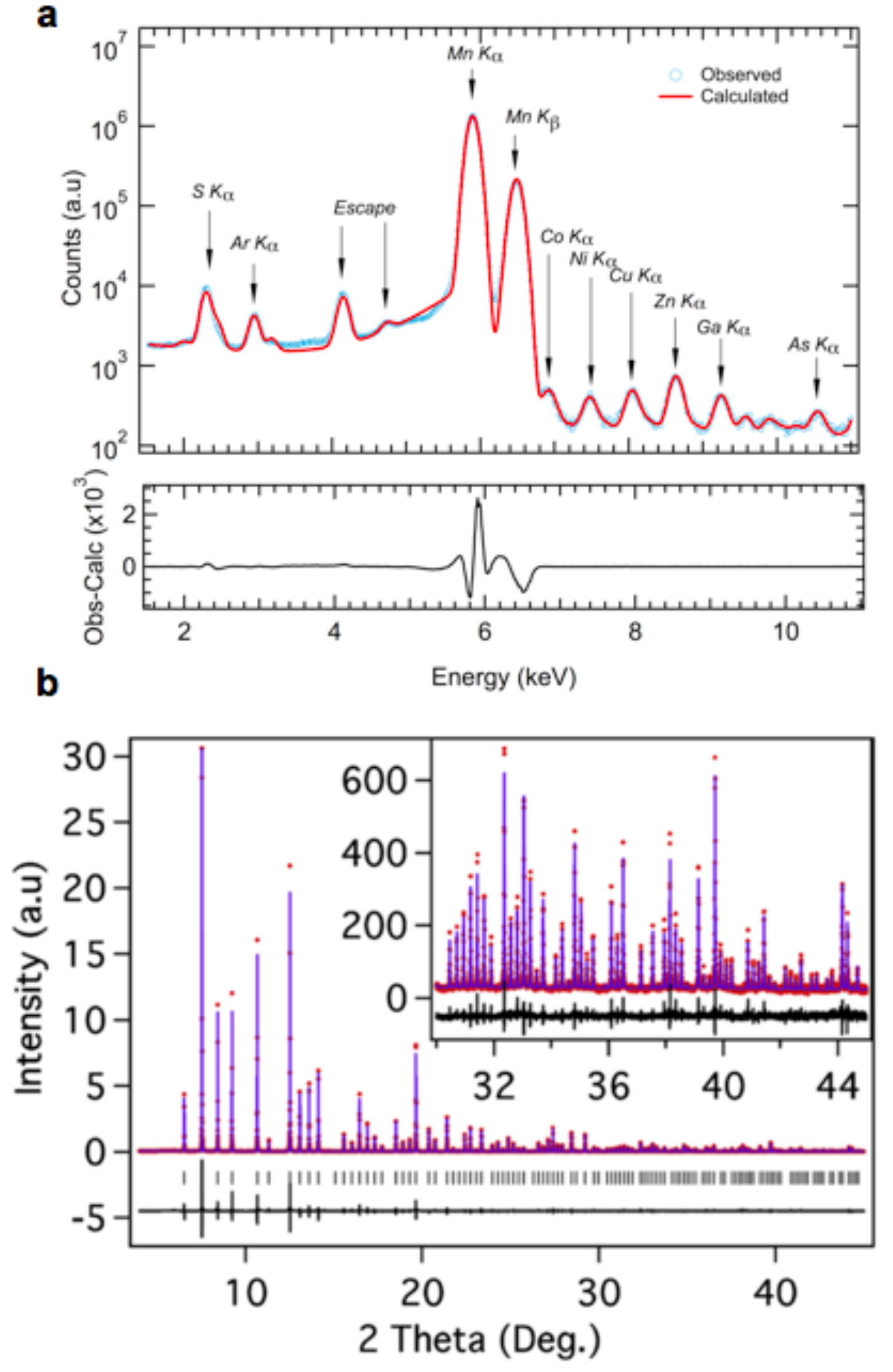}
\caption{\textbf{ S1: Characterisation of sample purity and ambient pressure structure.} \textbf{a}, X-ray fluorescence measurements showed the presence of minor transition metal impurities at a level of less than 0.1 \%. Similar levels of Se and As contamination were observed. The model calculation was performed using the PyMca software from the ESRF. \textbf{b}, Very high resolution powder x-ray diffraction data collected at 55 K on ID31 at the ESRF showed no deviation from the cubic \textit{Pa-3} pyrite structure, and no evidence for significant impurities. The refinement shown below converged with  $\chi^{2}$= 5.38 and gave a refined lattice parameter a = 6.088821(4) \AA~and a refined sulphur coordinate of x = 0.400061(6).  }
\label{Fig1}
\end{center}
\end{figure}

\begin{figure}
\begin{center}
\includegraphics[scale=0.65]{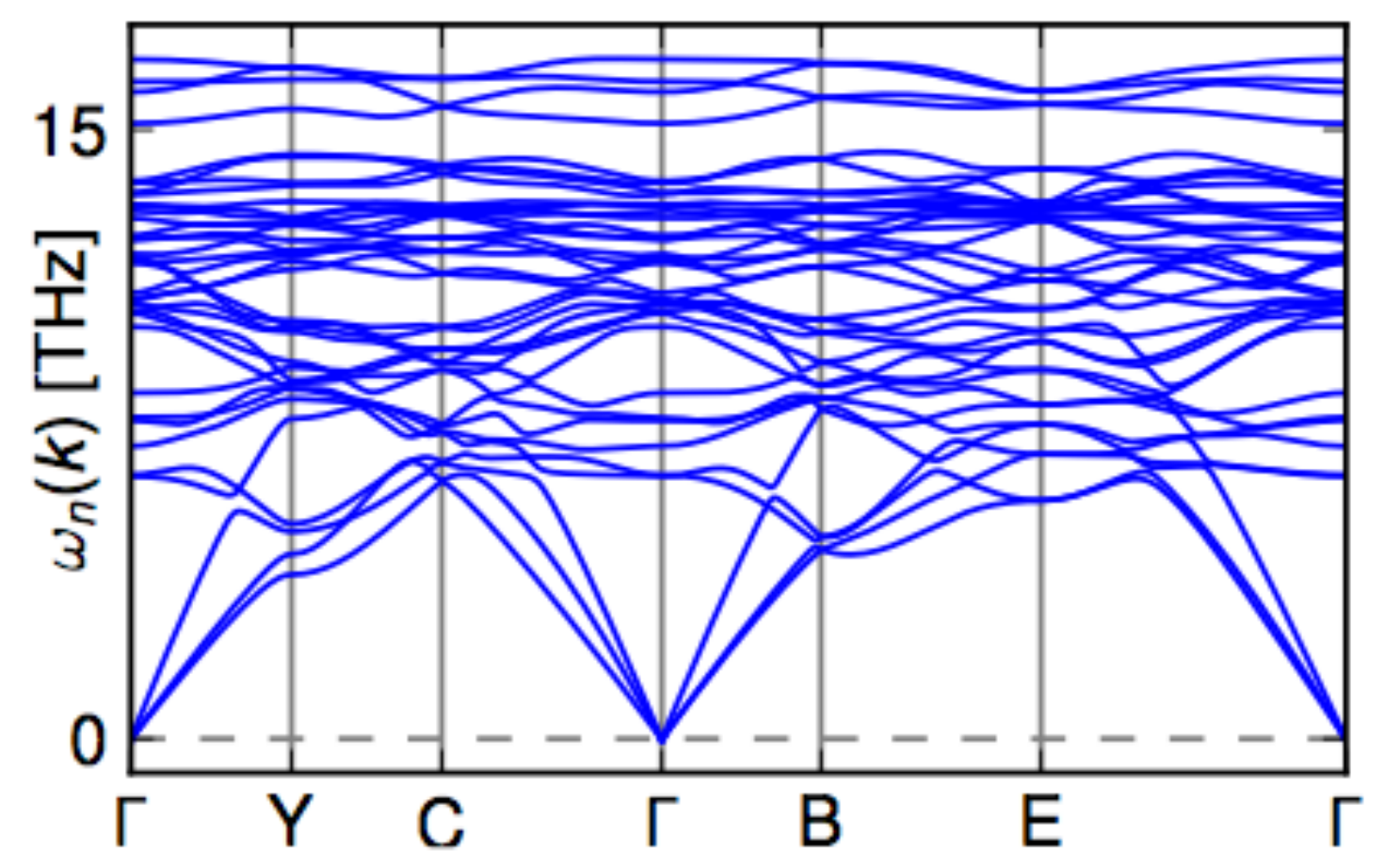}
\caption{\textbf{ S2: Phonon band structure of the high pressure polymorph of \mn at 20 GPa.} We investigated the stability of our high pressure structure by calculating the phonon band structure. All frequencies are real. The high symmetry points of the monoclinic $P\,2_1/c$ structure are $Y=(\frac{1}{2},0,0)$, $C=(\frac{1}{2},\frac{1}{2},0)$, $B=(0,0,\frac{1}{2})$, $E=(\frac{1}{2},\frac{1}{2},\frac{1}{2})$ }
\label{Fig1}
\end{center}
\end{figure}

\begin{figure}
\begin{center}
\includegraphics[scale=0.65]{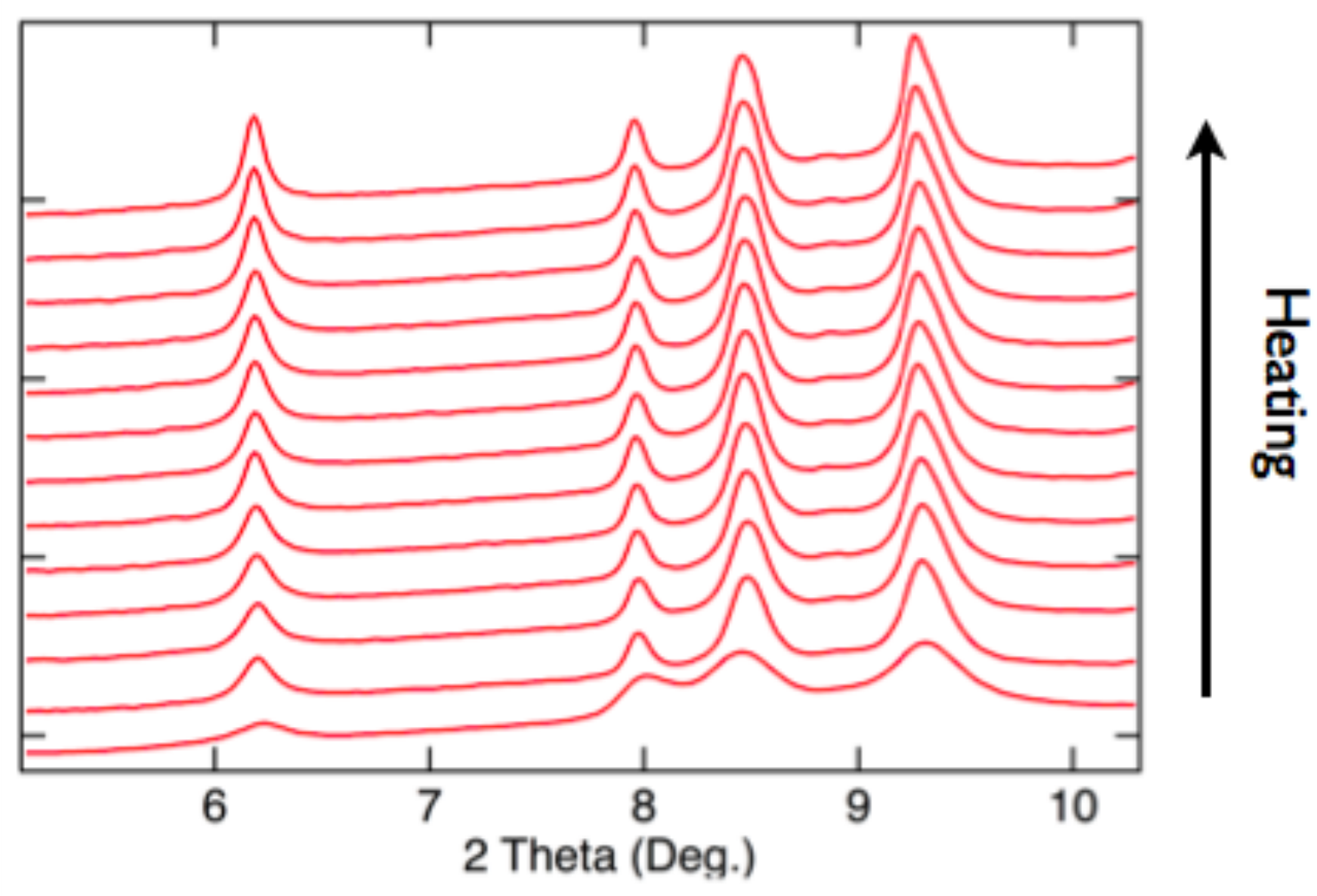}
\caption{\textbf{ S3: Diffraction data during \textit{in-situ}~laser-heating of \mn.} Data are shown immediately after commencing laser heating. The frame rate was approximately 1 Hz, and the highest temperature reached was $\sim$1800 K, as judged by thermal emission. }
\label{Fig1}
\end{center}
\end{figure}

\begin{figure}
\begin{center}
\includegraphics[scale=0.65]{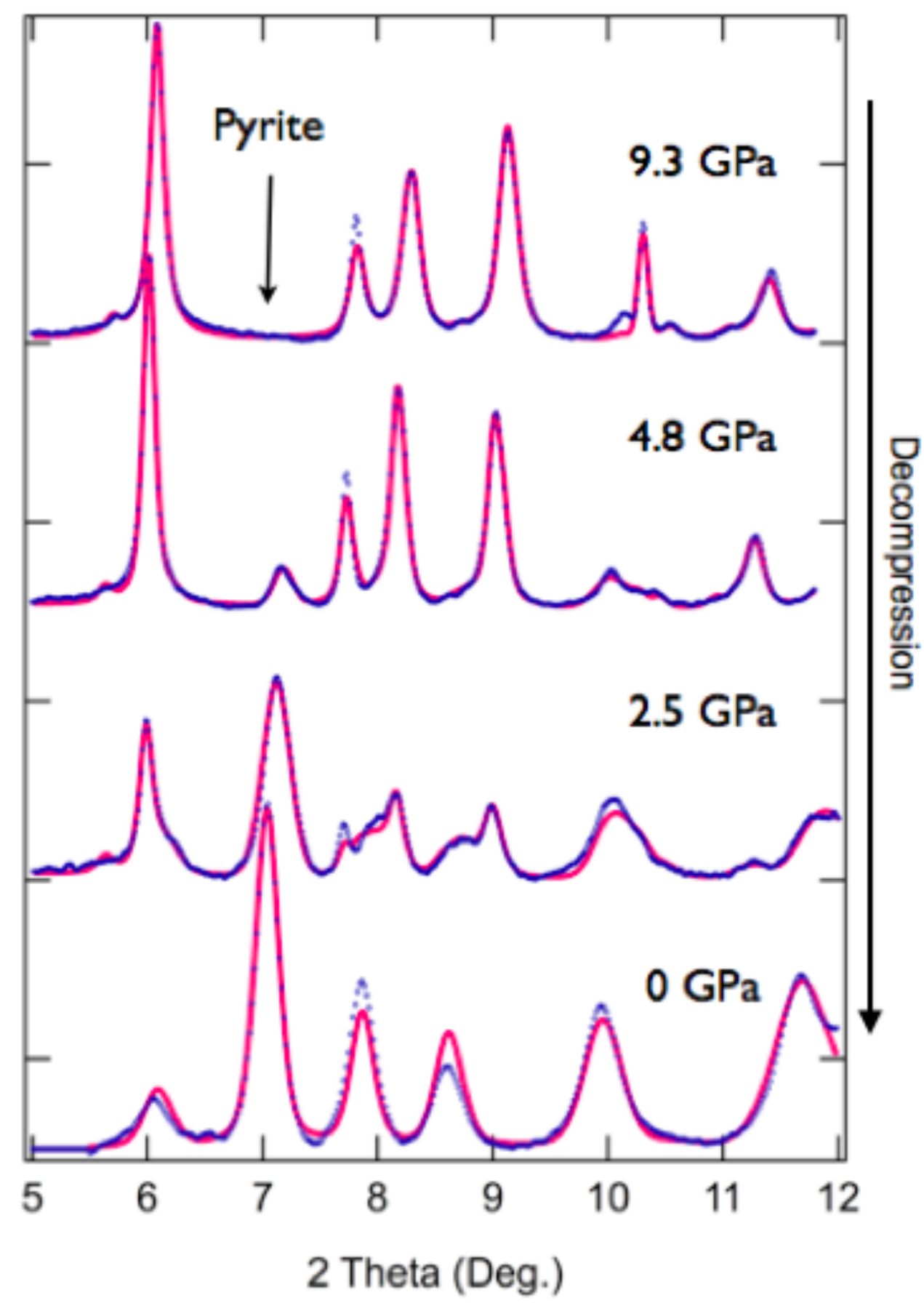}
\caption{\textbf{ S4: Diffraction data collected during the post-laser heating decompression of \mn.} The diffraction data shown below were collected on decompressing the sample which had been laser heated. The dots are data points and the lines are model calculations using the  arsenopyrite/pyrite structure models as appropriate. Note the onset of mixed phase behaviour below $\sim$ 5 GPa and the recovery of the pyrite phase. }
\label{Fig1}
\end{center}
\end{figure}

\begin{figure}
\begin{center}
\includegraphics[scale=0.4]{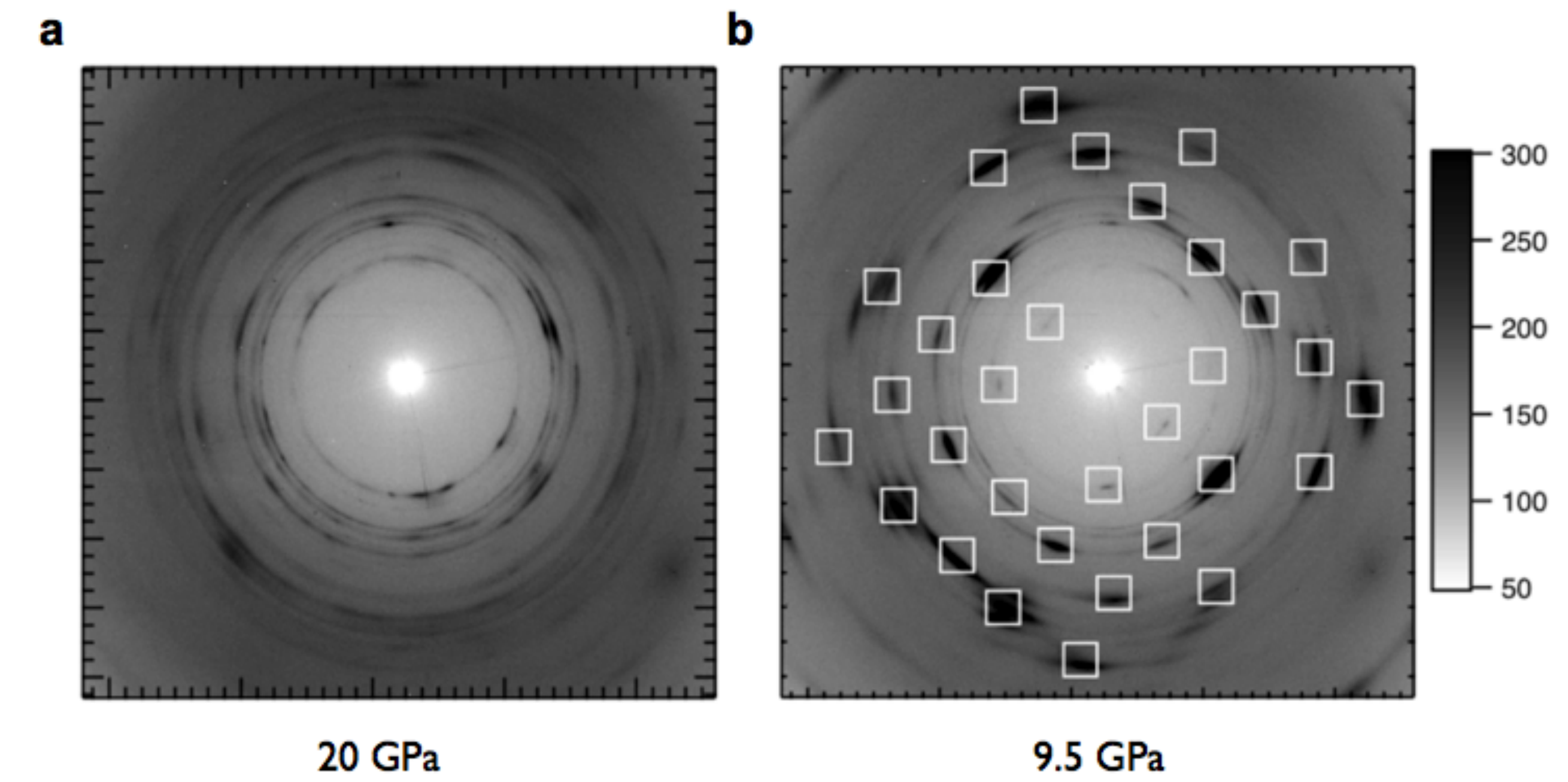}
\caption{\textbf{ S5: Texture effects on decompression} Raw diffraction images are shown for the ID9 data at the highest pressure (30 GPa), and after reducing pressure to 9.5 GPa. We discovered that strong texturing sets in on decompression as highlighted by the white squares in panel b.}
\label{Fig1}
\end{center}
\end{figure}

\begin{figure}
\begin{center}
\includegraphics[scale=0.4]{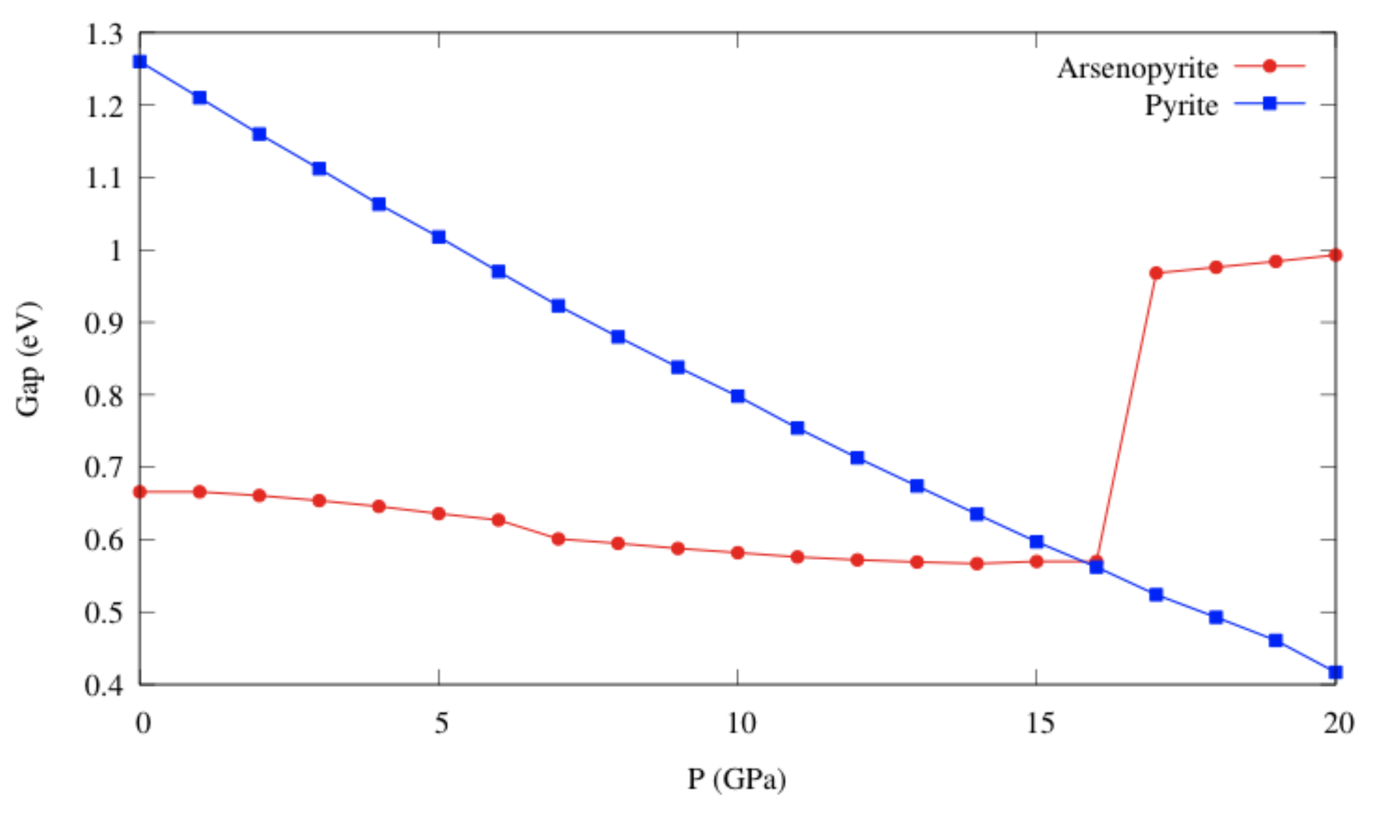}
\caption{\textbf{ S6: Pressure dependence of the band-gaps in pyrite and arsenopyrite \mn.} The pressure-dependence of the band gaps calculated using the \textsc{lapw}~method for both phases of \mn~is shown. Note the increase in the band gap of the arsenopyrite phase at 16 GPa, where the Mn-Mn dimerisation is enhanced.}
\label{Fig1}
\end{center}
\end{figure}

\begin{figure}
\begin{center}
\includegraphics[scale=0.45]{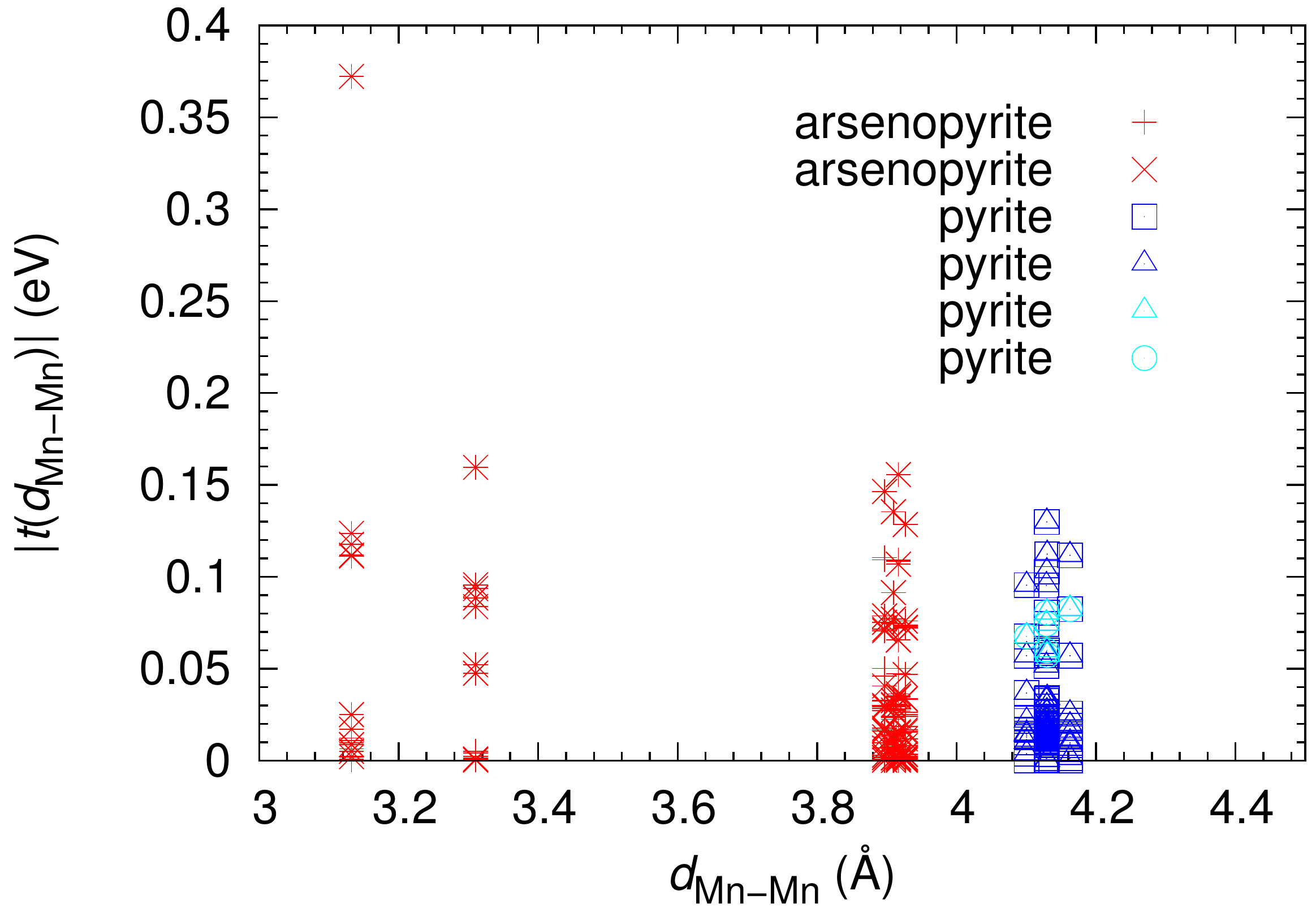}
\caption{\textbf{ S7: Results of tight-binding fits to the electronic structure of pyrite and arsenopyrite \mn.} Tight binding parameter values for the twenty Mn \textit{3}~bands in the pyrite and arsenopyrite structures at P=10 GPa, as obtained from projective Wannier functions within FPLO. One especially large hopping parameter for the shortest Mn-Mn bond appears in the arsenopyrite structure. It corresponds to a \textit{3d$_{xz}$}-\textit{3d$_{xz}$}~Wannier function overlap}
\label{Fig1}
\end{center}
\end{figure}

\begin{table*}
\caption{{\bf Calculated arsenopyrite structures of {\mn} from $P=10$~GPa to $P=20$~GPa.  Note that they were relaxed with Mn in type 3 antiferromagnetic order with GGA+U exchange and correlation functional at $U=3$~eV and $J=1$~eV.
}. }\label{tab:10}
\begin{ruledtabular}
\begin{tabular}{ccccccccc}
\multicolumn{9}{l}{$P=10$~GPa, space group $P\,2_1/c$, $a=5.59345$~{\AA}, $b=5.47203$~{\AA}, $c=5.58484$~{\AA}, $\beta=109.920^\circ$.}\\\hline
\multicolumn{1}{c}{\text{Mn }$x$}&\multicolumn{1}{c}{\text{Mn }$y$}&\multicolumn{1}{c}{\text{Mn }$z$}&\multicolumn{1}{c}{\text{S}$(1)\:x$}&\multicolumn{1}{c}{\text{S}$(1)\:y$}&\multicolumn{1}{c}{\text{S}$(1)\:z$}&\multicolumn{1}{c}{\text{S}$(2)\:x$}&\multicolumn{1}{c}{\text{S}$(2)\:y$}&\multicolumn{1}{c}{\text{S}$(2)\:z$}\\\hline
  -0.25112 &  -0.00017 &  -0.25197  &   -0.34402&   0.37298 &  -0.15687  &   -0.15568  & -0.37358  & -0.34563 \\\hline\hline
\multicolumn{9}{l}{$P=11$~GPa, space group $P\,2_1/c$, $a=5.57405$~{\AA}, $b=5.46472$~{\AA}, $c=5.57233$~{\AA}, $\beta=109.965^\circ$.}\\\hline
\multicolumn{1}{c}{\text{Mn }$x$}&\multicolumn{1}{c}{\text{Mn }$y$}&\multicolumn{1}{c}{\text{Mn }$z$}&\multicolumn{1}{c}{\text{S}$(1)\:x$}&\multicolumn{1}{c}{\text{S}$(1)\:y$}&\multicolumn{1}{c}{\text{S}$(1)\:z$}&\multicolumn{1}{c}{\text{S}$(2)\:x$}&\multicolumn{1}{c}{\text{S}$(2)\:y$}&\multicolumn{1}{c}{\text{S}$(2)\:z$}\\\hline
 -0.25316  & 0.49971 &  -0.25569&    -0.34380 &  -0.12766 &  -0.15920 &   -0.15530  & 0.12619 &  -0.34785\\\hline\hline
\multicolumn{9}{l}{$P=12$~GPa, space group $P\,2_1/c$, $a=5.55490$~{\AA}, $b=5.45567$~{\AA}, $c=5.56002$~{\AA}, $\beta=110.171^\circ$.}\\\hline
\multicolumn{1}{c}{\text{Mn }$x$}&\multicolumn{1}{c}{\text{Mn }$y$}&\multicolumn{1}{c}{\text{Mn }$z$}&\multicolumn{1}{c}{\text{S}$(1)\:x$}&\multicolumn{1}{c}{\text{S}$(1)\:y$}&\multicolumn{1}{c}{\text{S}$(1)\:z$}&\multicolumn{1}{c}{\text{S}$(2)\:x$}&\multicolumn{1}{c}{\text{S}$(2)\:y$}&\multicolumn{1}{c}{\text{S}$(2)\:z$}\\\hline
  -0.25700 &  0.49947  & -0.26255  &    -0.34304 &  -0.12850 &  -0.16333  &  -0.15490 &  0.12538 &  -0.35198 
\\\hline\hline
\multicolumn{9}{l}{$P=13$~GPa, space group $P\,2_1/c$, $a=5.54785$~{\AA}, $b=5.45037$~{\AA}, $c=5.55599$~{\AA}, $\beta=109.960^\circ$.}\\\hline
\multicolumn{1}{c}{\text{Mn }$x$}&\multicolumn{1}{c}{\text{Mn }$y$}&\multicolumn{1}{c}{\text{Mn }$z$}&\multicolumn{1}{c}{\text{S}$(1)\:x$}&\multicolumn{1}{c}{\text{S}$(1)\:y$}&\multicolumn{1}{c}{\text{S}$(1)\:z$}&\multicolumn{1}{c}{\text{S}$(2)\:x$}&\multicolumn{1}{c}{\text{S}$(2)\:y$}&\multicolumn{1}{c}{\text{S}$(2)\:z$}\\\hline
  -0.25616 &  -0.00053  & -0.26113   &    -0.34320 &  0.37155  & -0.16257   &    -0.15492 &  -0.37429 &  -0.35132
\\\hline\hline
\multicolumn{9}{l}{$P=14$~GPa, space group $P\,2_1/c$, $a=5.53365$~{\AA}, $b=5.44896$~{\AA}, $c=5.54880$~{\AA}, $\beta=110.396^\circ$.}\\\hline
\multicolumn{1}{c}{\text{Mn }$x$}&\multicolumn{1}{c}{\text{Mn }$y$}&\multicolumn{1}{c}{\text{Mn }$z$}&\multicolumn{1}{c}{\text{S}$(1)\:x$}&\multicolumn{1}{c}{\text{S}$(1)\:y$}&\multicolumn{1}{c}{\text{S}$(1)\:z$}&\multicolumn{1}{c}{\text{S}$(2)\:x$}&\multicolumn{1}{c}{\text{S}$(2)\:y$}&\multicolumn{1}{c}{\text{S}$(2)\:z$}\\\hline
 -0.25912 &  0.49923 &  -0.26632   &    -0.34242 &  -0.12912 &  -0.16642  &   -0.15494 &  0.12537  & -0.35440
\\\hline\hline
\multicolumn{9}{l}{$P=15$~GPa, space group $P\,2_1/c$, $a=5.46505$~{\AA}, $b=5.42287$~{\AA}, $c=5.50583$~{\AA}, $\beta=111.581^\circ$.}\\\hline
\multicolumn{1}{c}{\text{Mn }$x$}&\multicolumn{1}{c}{\text{Mn }$y$}&\multicolumn{1}{c}{\text{Mn }$z$}&\multicolumn{1}{c}{\text{S}$(1)\:x$}&\multicolumn{1}{c}{\text{S}$(1)\:y$}&\multicolumn{1}{c}{\text{S}$(1)\:z$}&\multicolumn{1}{c}{\text{S}$(2)\:x$}&\multicolumn{1}{c}{\text{S}$(2)\:y$}&\multicolumn{1}{c}{\text{S}$(2)\:z$}\\\hline
  -0.27172 &  0.00018 &  -0.28596  &   -0.34104 &  0.36922 &  -0.17620  &    -0.15378 &  -0.37473 &  -0.36770 
\\\hline\hline
\multicolumn{9}{l}{$P=16$~GPa, space group $P\,2_1/c$, $a=5.46440$~{\AA}, $b=5.41649$~{\AA}, $c=5.49656$~{\AA}, $\beta=111.442^\circ$.}\\\hline
\multicolumn{1}{c}{\text{Mn }$x$}&\multicolumn{1}{c}{\text{Mn }$y$}&\multicolumn{1}{c}{\text{Mn }$z$}&\multicolumn{1}{c}{\text{S}$(1)\:x$}&\multicolumn{1}{c}{\text{S}$(1)\:y$}&\multicolumn{1}{c}{\text{S}$(1)\:z$}&\multicolumn{1}{c}{\text{S}$(2)\:x$}&\multicolumn{1}{c}{\text{S}$(2)\:y$}&\multicolumn{1}{c}{\text{S}$(2)\:z$}\\\hline
   -0.27002 &   0.00028 &   -0.28314 &  -0.34156 &   0.36888  &  -0.17483    &   -0.15350  &  -0.37510  &  -0.36454 
\\\hline\hline
\multicolumn{9}{l}{$P=17$~GPa, space group $P\,2_1/c$, $a=5.43310$~{\AA}, $b=5.40310$~{\AA}, $c=5.48509$~{\AA}, $\beta= 111.833^\circ$.}\\\hline
\multicolumn{1}{c}{\text{Mn }$x$}&\multicolumn{1}{c}{\text{Mn }$y$}&\multicolumn{1}{c}{\text{Mn }$z$}&\multicolumn{1}{c}{\text{S}$(1)\:x$}&\multicolumn{1}{c}{\text{S}$(1)\:y$}&\multicolumn{1}{c}{\text{S}$(1)\:z$}&\multicolumn{1}{c}{\text{S}$(2)\:x$}&\multicolumn{1}{c}{\text{S}$(2)\:y$}&\multicolumn{1}{c}{\text{S}$(2)\:z$}\\\hline
  -0.27380 &  0.00029 &  -0.28821   &   -0.34148 &  0.36845  & -0.17739   &    -0.15302 &  -0.37532  & -0.36762 
\\\hline\hline
\multicolumn{9}{l}{$P=18$~GPa, space group $P\,2_1/c$, $a=5.42415$~{\AA}, $b=5.39539$~{\AA}, $c=5.47696$~{\AA}, $\beta=111.869^\circ$.}\\\hline
\multicolumn{1}{c}{\text{Mn }$x$}&\multicolumn{1}{c}{\text{Mn }$y$}&\multicolumn{1}{c}{\text{Mn }$z$}&\multicolumn{1}{c}{\text{S}$(1)\:x$}&\multicolumn{1}{c}{\text{S}$(1)\:y$}&\multicolumn{1}{c}{\text{S}$(1)\:z$}&\multicolumn{1}{c}{\text{S}$(2)\:x$}&\multicolumn{1}{c}{\text{S}$(2)\:y$}&\multicolumn{1}{c}{\text{S}$(2)\:z$}\\\hline
    -0.27388 &  -0.49972 &  -0.28839  &     -0.34134 &  -0.13157 &  -0.17760  &   -0.15306 &  0.12472 &  -0.36771 
\\\hline\hline
\multicolumn{9}{l}{$P=19$~GPa, space group $P\,2_1/c$, $a=5.41550$~{\AA}, $b=5.38817$~{\AA}, $c=5.46917$~{\AA}, $\beta=111.888^\circ$.}\\\hline
\multicolumn{1}{c}{\text{Mn }$x$}&\multicolumn{1}{c}{\text{Mn }$y$}&\multicolumn{1}{c}{\text{Mn }$z$}&\multicolumn{1}{c}{\text{S}$(1)\:x$}&\multicolumn{1}{c}{\text{S}$(1)\:y$}&\multicolumn{1}{c}{\text{S}$(1)\:z$}&\multicolumn{1}{c}{\text{S}$(2)\:x$}&\multicolumn{1}{c}{\text{S}$(2)\:y$}&\multicolumn{1}{c}{\text{S}$(2)\:z$}\\\hline
  -0.27380 &  0.00024  & -0.28843  &   -0.34118  & 0.36844 &  -0.17775   &    -0.15312  & -0.37524 &  -0.36779  
\\\hline\hline
\multicolumn{9}{l}{$P=20$~GPa, space group $P\,2_1/c$, $a=5.40735$~{\AA}, $b=5.38104$~{\AA}, $c=5.46141$~{\AA}, $\beta=111.912^\circ$.}\\\hline
\multicolumn{1}{c}{\text{Mn }$x$}&\multicolumn{1}{c}{\text{Mn }$y$}&\multicolumn{1}{c}{\text{Mn }$z$}&\multicolumn{1}{c}{\text{S}$(1)\:x$}&\multicolumn{1}{c}{\text{S}$(1)\:y$}&\multicolumn{1}{c}{\text{S}$(1)\:z$}&\multicolumn{1}{c}{\text{S}$(2)\:x$}&\multicolumn{1}{c}{\text{S}$(2)\:y$}&\multicolumn{1}{c}{\text{S}$(2)\:z$}\\\hline
   -0.27380 &  0.00020  & -0.28848 &   -0.34108 &  0.36845  & -0.17792   &   -0.15312 &  -0.37522 &  -0.36777 
\end{tabular}
\end{ruledtabular}
\end{table*}

\begin{table*}
\caption{Refined  pyrite and arsenopyrite structures of {\mn} at P= 0 and 20~GPa respectively. In the pyrite structure, the Mn atoms are on the \textit{4a}~position, and the S atoms are on the \textit{8c}~position. The pyrite density is 4.85 g/cm$^{3}$.  In the arsenopyrite structure, the Mn atoms, and the two independent S sites are on the \textit{4e}~position. The arsenopyrite density is 6.385 g/cm$^{3}$.}\label{tab:0}
\begin{ruledtabular}
\begin{tabular}{ccccccccc}
\multicolumn{9}{l}{$P=0$~GPa, space group $Pa-3$, $a=6.0888214(5)$, S($x$)=0.400061(6)~{\AA}.}\\\hline
\multicolumn{9}{l}{$P=20$~GPa, space group $P\,2_1/c$, $a=5.4642(6)$~{\AA}, $b=5.3742(5)$~{\AA}, $c=5.4203(7)$~{\AA}, $\beta=111.93(2)^\circ$.}\\\hline
\multicolumn{1}{c}{\text{Mn }$x$}&\multicolumn{1}{c}{\text{Mn }$y$}&\multicolumn{1}{c}{\text{Mn }$z$}&\multicolumn{1}{c}{\text{S}$(1)\:x$}&\multicolumn{1}{c}{\text{S}$(1)\:y$}&\multicolumn{1}{c}{\text{S}$(1)\:z$}&\multicolumn{1}{c}{\text{S}$(2)\:x$}&\multicolumn{1}{c}{\text{S}$(2)\:y$}&\multicolumn{1}{c}{\text{S}$(2)\:z$}\\\hline
 0.280(2)  & 0.012(2) &  0.288(1)&    0.345(2)&  0.375(2) &  0.175(2) &   -0.139(2)  & 0.601(2) &  -0.368(2)
\\\hline
\end{tabular}
\end{ruledtabular}
\end{table*}

\clearpage

\end{document}